\documentclass[12pt,onecolumn]{IEEEtran}
\usepackage[colorlinks,bookmarksopen,bookmarksnumbered,citecolor=red,urlcolor=red,]{hyperref}
 \usepackage{CJK}
 \usepackage{indentfirst}
\usepackage{algorithm,algorithmic,amsbsy,amsmath,amssymb,epsfig,bbm,mathrsfs, bbm} 
\usepackage{multirow}
\usepackage{mathrsfs}
 \usepackage{epstopdf}

\usepackage{verbatim}
\usepackage{amsfonts}
\usepackage{amsthm}
\usepackage[subfigure]{graphfig}

\begin{document}

\title{Dynamic Spectrum Access in Cognitive Radio Networks with RF Energy Harvesting}
\author{Lu Xiao$^1$, Ping Wang$^1$, Dusit Niyato$^1$, and Ekram Hossain$^2$	\\
$^1$ School of Computer Engineering, Nanyang Technological University (NTU), Singapore\\
$^2$ Electrical and Computer Engineering, University of Manitoba, Canada \vspace{-5mm}
\thanks{\textbf{D. Niyato} is the corresponding author.}}

\maketitle

\begin{abstract}
Spectrum efficiency and energy efficiency  are two critical issues in designing wireless networks.  Through dynamic spectrum access, cognitive radios can improve the spectrum efficiency and capacity of wireless networks.  On the other hand, radio frequency (RF) energy harvesting has emerged as a promising technique to supply energy to wireless networks and thereby increase their energy efficiency. Therefore, to achieve both spectrum and energy efficiencies, the secondary users in a cognitive radio network (CRN) can be equipped with the RF energy harvesting capability and such a network can be referred to as an RF-powered cognitive radio network. In this article, we provide  an overview of the RF-powered CRNs and discuss the challenges that arise for dynamic spectrum access in these networks. Focusing on the tradeoff among spectrum sensing, data transmission, and RF energy harvesting, then we discuss the dynamic channel selection problem in a multi-channel RF-powered CRN. In the RF-powered CRN, a secondary user can adaptively select a channel to transmit data when the channel is not occupied by any primary user. Alternatively, the secondary user can harvest RF energy for data transmission if the channel is occupied. The optimal channel selection policy of the secondary user can be obtained by formulating a Markov decision process (MDP) problem. We present some numerical results obtained by solving this MDP problem.
\end{abstract}

\section{Introduction}

Energy supply is always a critical issue in wireless communications~\cite{Hasan2011Ziaul}. Traditionally, portable/mobile wireless nodes operate with energy supply from a battery, which has a limited capacity and needs to be physically charged or replaced regularly. Recently, RF energy harvesting technology has been developed and is able to supply energy to wireless nodes. Table~\ref{Data} summarizes the experimental measurement of RF energy harvested from various RF energy sources. Besides, an important study~\cite{Chomora2011} on the design of a digital TV energy harvesting circuit reports RF-to-DC conversion efficiencies above $0.4\%$ at - 40 dBm, above $18.2\%$ at -20dBm and over $50\%$ at -5 dBm RF signal power incidence. Both the sensitivity of energy harvester and the conversion efficiency are expected to be improved in the near future. Therefore, the adoption of RF energy harvesting technology is very plausible. In addition, compared with other forms of energy harvesting (e.g., solar, vibration, wind and acoustic noise), RF energy harvesting does not depend on the Nature, and hence it provides relatively predictable energy supply. The amount of RF energy that can be harvested depends on the wavelength of the harvested RF signal and the distance between an RF energy source and the harvesting device (Table I), which can be calculated based on the Friis transmission equation~\cite{Cuadras2010}. 

\begin{table}\footnotesize
\centering
\caption{\footnotesize Experimental data of RF Energy Harvesting.} \label{Data}
\begin{tabular}{|l|l|l|l|l|  }
\hline
\footnotesize Source & Source Power & Frequency & Distance & Amount of Energy Harvested     \\ \hline
\hline
GSM900 \cite{Visser2013}& & 935-960MHZ & 25m-100m & $10^{-3}-10^{-1}\mu$ W/cm$^{2}$    \\
\hline
GSM1800 \cite{Visser2013} & & 1805.2-1879.8MHZ  & 25m-100m &  $10^{-3}-10^{-1}\mu$ W/cm$^{2}$   \\
\hline
AM Radio Station \cite{Ostaffe2010} & 50KW &    &  5KM & 159$\mu$ W/m$^{2}$   \\
 & 50KW &     &  10KM & 40$\mu$ W/m$^{2}$  \\
 & 5KW  &     & 2.5KM & 200$\mu$W   \\
\hline
Mobile Base Station \cite{Ostaffe2010}  & 100W & & 100m & 800$\mu$W/m$^{2}$   \\
& 100W  & & 500m &  32$\mu$W/m$^{2}$   \\
& 100W  & & 1000m &  8$\mu$ W/m$^{2}$   \\
\hline
Mobile Phone \cite{Ostaffe2010} & 0.5W & 915 MHz & $1m$ & $40mW/m^{2}$    \\
& 0.5W &   915MHz & 5m &  1.6mW/$m^{2}$    \\
& 0.5W  & 915MHz  &10m  & 0.4mW/m$^{2}$    \\
\hline
TX91501 Powercaster Transmitter \cite{Murtala2012}  & 3W & 915MHZ & $5m$ & 250$\mu$ W$/cm^{2}$   \\
\hline
\end{tabular}
\end{table}

Powering a cognitive radio network (CRN) with RF energy can provide a spectrum-efficient and energy-efficient solution for wireless networking~\cite{SPark2013,SLee2013}. In an RF-powered cognitive radio network (RF-powered CRN), the RF energy harvesting capability allows the wireless devices (e.g., secondary users) to harvest energy from RF signals and use that energy for their data transmission. Such RF signals could be from nearby RF sources (e.g., primary users, cellular base stations, and other ambient RF sources) and can be converted into DC electricity. This energy can be stored in an energy storage and used to operate the devices and transmit data. To save cost and reduce implementation complexity, the wireless interface of the cognitive radio devices in an RF-powered CRN can be reused for RF energy harvesting in addition to transmitting and receiving data. 
The secondary users can transmit data when they are sufficiently far away from primary users or when the nearby primary users are idle. Therefore, the devices must not only identify spectrum  holes for opportunistic data transmission, but also search for occupied spectrum band to harvest RF energy.\footnote{In this paper, we use the terms ``channel" and ``spectrum band" interchangeably.} Due to specific nature of RF energy harvesting (e.g., the amount of harvested RF energy depends on distance) and the communication requirements of the cognitive radio devices, the communication protocols for the traditional CRNs may not be efficient for RF-powered CRNs. In particular, the dynamic spectrum sensing and channel access methods for the cognitive radios have to be optimized considering the tradeoff among network throughput (or spectrum efficiency), energy efficiency, and RF energy supply. 

Recent literature on RF-powered CRNs mainly focuses on investigating throughput maximization under various constraints. For example, \cite{SPark2012} considers RF energy harvesting-enabled cognitive radio sensor networks under an energy causality constraint. The constraint imposes that the total consumed energy should not be greater than the total harvested energy. The authors propose an optimal mode selection policy to balance between the immediate throughput and harvested RF energy in transmitting and harvesting modes, respectively. In~\cite{SLee2013}, the mobile devices in a secondary network opportunistically either harvest RF energy from  transmissions of nearby devices in a primary network, or transmit data if the devices are not in the interference range of any other primary network. The throughput of the secondary network is maximized by deriving the optimal transmit power and density of the secondary transmitters under an outage-probability constraint. In~\cite{Norberto2012Barroca}, the authors consider a cognitive wireless body area network with RF energy harvesting capability. The authors discuss the challenges in the physical, medium access control (MAC), and network layers and  some potential solutions. In addition, practical architectures are proposed for cognitive radio-enabled RF energy harvesting devices for joint information reception and RF energy harvesting. However, the problem of dynamic spectrum access for RF-powered CRNs has not been rigorously studied in the literature, and this is the main focus of this article.

We first present an overview of the RF-powered CRNs and highlight the main differences between traditional CRN and RF-powered CRNs. Then we discuss the research challenges related to dynamic spectrum access in the RF-powered CRNs. Then, to study the tradeoff between spectrum sensing, data transmission, and RF energy harvesting, we focus on the problem of channel selection for dynamic spectrum access in a multi-channel RF-powered CRN. This RF-powered CRN consists of multiple primary users allocated with different channels and secondary users with RF energy harvesting capability. The objective of a secondary user is to maximize her throughput. To achieve this objective, a channel selection policy has to be used. This policy is a mapping of the secondary user's state (i.e., data queue, energy storage, and channel status) to a particular channel to sense and transmit data or harvest RF energy. The policy can be obtained by formulating an optimization problem based on the Markov decision process.



\section{Overview of RF-Powered Cognitive Radio Network}
\label{sec:overview}

\subsection{RF Energy Harvesting in Cognitive Radio Device}

\begin{figure}
\centering
\includegraphics[width=0.5\textwidth]{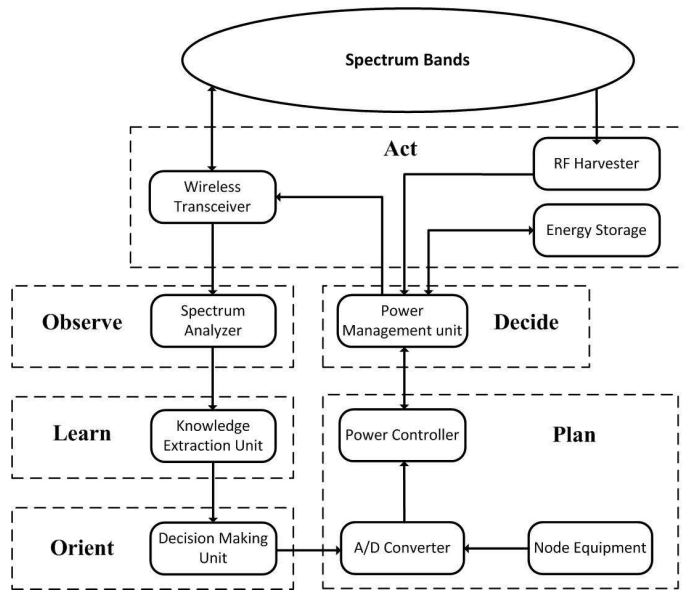}
\caption{Components in RF-powered CRN device and their relationship to cognition cycle.} 
\label{component}
\end{figure}

Figure~\ref{component} shows the general block diagram of a cognitive radio device with RF energy harvesting. The device consists of following components.
\begin{itemize}
	\item A software-defined radio-based wireless transceiver for data transmission and reception,
	\item A spectrum analyzer which observes and analyzes the activity of spectrum usage from measured signals,
	\item A knowledge extraction unit which uses the information on spectrum usage to build and maintain a knowledge base of the spectrum access environment,
	\item A decision making unit to make decision on spectrum access based on the knowledge base,
	\item A node equipment which implements certain applications (e.g., sensors),
	\item An A/D converter that digitizes the analog signal from the node equipment,
	\item A power controller to process the digital signal from A/D converter for network applications,
	\item An energy storage device which could be a battery or capacitor (for a low-power node) to store the harvested energy for future use,
	\item A power management unit, which dispatches the energy from RF energy harvester (i.e., decide whether to store the harvested energy in a battery or to transfer it immediately to other components), and
	\item An RF energy harvester to collect RF signal and covert it into electricity.
\end{itemize}
For a cognitive radio device, the major functions of  observing, learning, orienting, planning, deciding and acting can be represented as a cognition cycle~\cite{Fette2006} as shown in Fig.~\ref{component}.  

\begin{figure}
\centering
\includegraphics[width=0.8\textwidth]{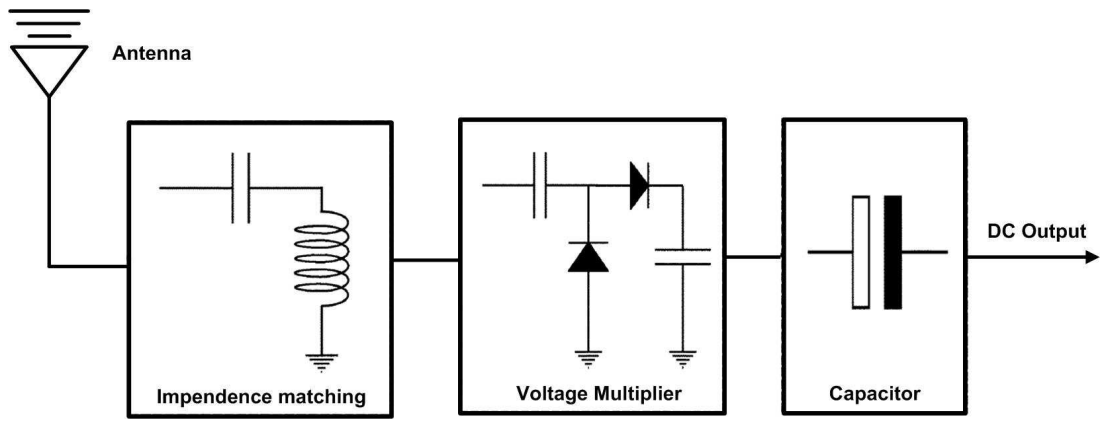}
\caption{Circuit design of an RF harvester.} 
\label{harvester}
\end{figure}

Figure \ref{harvester} shows the block diagram of a typical circuit  for an RF harvester which consists of antenna, impedance matching unit, voltage multiplier, and capacitor. 
\begin{itemize}
\item The impedance matching unit is a resonator circuit which operates at a designated frequency to maximize the power transfer between the antenna and the multiplier.
\item The main components of the voltage multiplier are the diodes of the rectifying circuit which converts RF waves (AC signal in nature) into DC signal. Generally, a higher conversion efficiency can be achieved by diodes with lower built-in voltage. 
\item The capacitor ensures a smooth delivery of power to the load. When energy harvesting is unavailable, the capacitor can also temporarily serve as a small energy reservoir. 
\end{itemize}

The RF harvester can be designed to work on either single frequency or multiple frequencies concurrently \cite{SKeyrouz2013}. 
The antenna may need to work at multiple frequencies simultaneously to acquire enough energy as input. 
Note that the wireless transceiver and RF energy harvester may use different wireless interfaces (e.g., antenna) or the same interface. For the former, the device can transmit and receive data and harvest RF energy at the same time, if they use different frequency. For the latter, the device will not be able to harvest RF energy at the same time as transmitting data. However, the device may be able to receive data and harvest RF energy simultaneously~\cite{RZhang2013}. 

\subsection{Architecture of RF-Powered Cognitive Radio Network}

\begin{figure}
\centering
\includegraphics[width=0.7\textwidth]{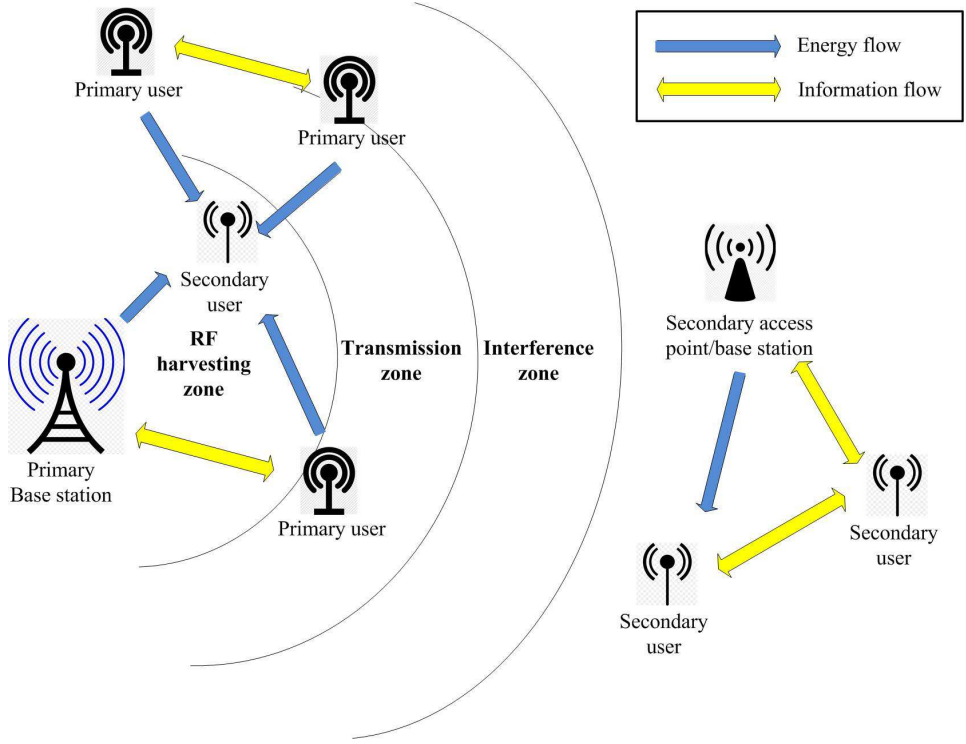}
\caption{Network architecture of RF-powered cognitive radio networks.} \label{architecture}
\end{figure}

An RF-powered CRN can be in various forms such as a cognitive wireless sensor network, a cognitive cellular network, a cognitive wireless mesh network, a cognitive device-to-device network, a cognitive wireless local area network, a cognitive wireless body network, etc. Either spectrum overlay or spectrum underlay can be adopted for the spectrum access of secondary users according to the used radio transmission technology and/or the network requirement. Figure~\ref{architecture} shows the general network architecture. In this architecture, RF signal is used not only to transmit data, but also to transfer energy. The secondary user can receive RF energy from the primary base station and other primary users. Alternatively, the secondary user can receive RF energy from a secondary base station and other secondary users. Figure~\ref{architecture} also shows three zones associated with the primary base station. The ``transmission zone'' is the coverage of the primary base station (e.g., a cell), where the primary user communicates with the primary base station. In side the ``transmission zone'', if the secondary user is in the ``RF harvesting zone'', the secondary user can harvest RF energy from the primary base station due to strong primary RF signal. If the primary base station or primary users occupy the spectrum, then the secondary user cannot transmit data if it is in the ``interference zone'' (i.e., interference is created to the communication of the primary users).

Like the conventional CRNs, the RF-powered CRNs can adopt either an infrastructure-based or an infrastructure-less communication architecture. In the infrastructure-based architecture, a secondary base station coordinates data communication among secondary users. Again, either centralized or distributed  dynamic spectrum access architecture~\cite{EHossain2009} can be used for different forms of RF-powered CRNs. In the former case, an optimal control of spectrum access can be achieved based on the global information about the radio environment and available RF energy gathered by a secondary base station. In the latter case, network-wise optimal control may not be achieved as the decisions on spectrum access and RF energy harvesting are made by individual secondary users autonomously and independently based on local information.

To optimize the performance of RF-powered CRNs, spectrum sensing, access, and handoff functionalities must be revisited. In the next section, we discuss about the issues of designing dynamic spectrum access for the RF-powered CRNs.


\section{Research Challenges in Dynamic Spectrum Access in RF-powered Cognitive Radio Networks}
\label{sec:issues}

To support intelligent and efficient dynamic spectrum access, cognitive radio networks  have four main functionalities, namely, spectrum sensing, spectrum access, spectrum management, and spectrum handoff.  In RF-powered CRNs, the cognitive radios have to identify and switch to not only idle channels for data transmission, but also sense the occupied channels for RF energy harvesting. Therefore, the traditional methods for spectrum sensing and access may not be sufficient for RF-powered CRNs.  In this section, we discuss the research issues that arise for spectrum sensing and access in RF-powered CRNs.

\subsection{Spectrum Sensing}
 
In conventional CRNs, a secondary user focuses on identifying spectrum holes, channel idle probability and channel quality. Some of the channel selection schemes, such as those in~\cite{Zhao2008, El-Sherif2011}, utilize the statistical spectrum occupancy information obtained through spectrum sensing to opportunistically access a free channel. However, the secondary user may select the channel that has a high idle probability but low channel quality. Therefore, channel selection schemes can also take channel quality into account~\cite{Cheung2011}. 


In the multi-channel RF-powered CRNs, in addition to finding a free channel and its quality, the secondary user has to identify an occupied channel and its RF signal whose energy can be harvested. To maximize the throughput, the secondary user will have the following preferences in channel selection. If the secondary user has a low energy level in its energy storage, it should select the channel which tends to be occupied by a primary user and has strong RF signal to harvest energy. On the other hand, if a secondary user has a high energy level and there are many packets waiting for transmission, it should select the channel which is likely to be idle with high probability of successful packet transmission (i.e., channel quality is favorable). In addition, the secondary user may decide not to transmit in some particular channels if the primary users are sensitive to interference. Clearly, the decision problem for the secondary users is more complex and hence the traditional channel selection schemes will not be sufficient. Therefore, channel selection policies specifically designed for the multi-channel RF-powered CRNs will be needed. 
 
Again, secondary users perform spectrum sensing periodically, where the duration and the frequency of sensing can be adjusted. A longer sensing duration and/or a higher sensing frequency will lead to a higher sensing accuracy and more harvested energy. However, the throughput performance will be adversely affected since there will be less time for the secondary user to transmit data. Thus there exists a tradeoff among the sensing duration and frequency of sensing (and hence sensing accuracy) and the amount of harvested energy versus communication throughput. The spectrum sensing period and the frequency of sensing can be optimized by jointly considering this tradeoff. 

The secondary users can use either a proactive or an on-demand approach for spectrum sensing. In the proactive approach, a secondary user periodically senses different channels and maintains a database of all the channels. In the on-demand approach, the secondary user may sense the target channel when it needs to switch to the new channel. Another related issue would be the order in which  the channels need to be sensed and selection of the channel to switch to. This decision depends on the activities of the primary users and also the state of secondary users (e.g., remaining energy level).


\subsection{Spectrum Access}

Traditionally, the spectrum access or MAC protocols for spectrum overlay-based CRNs are designed with the objective of maximizing the throughput of secondary users while protecting primary user from collisions due to secondary transmissions  and to provide fair and efficient sharing of available spectrum among secondary users. For the RF-powered CRNs, two types of MAC protocols, i.e., fixed and random spectrum access, can be adopted to achieve similar performance objectives.  

\begin{itemize}
	\item {\em Fixed spectrum access}: For this type of protocols, radio resources (time slots and channels or subcarriers) are statically allocated to users (e.g., based on time-division multiple access [TDMA], or orthogonal frequency-division multiple access [OFDMA]). Given the availability of RF energy, the radio resource must be allocated optimally among multiple secondary users. For example, radio resource should be allocated to the users which are not harvesting RF energy (e.g., out of range of transmitting RF sources). Also, the radio resource should be allocated to the users which have sufficient amount of harvested RF energy to use the allocated radio bandwidth. 
	\item  {\em Random spectrum access}: For this type of protocols (e.g., slotted ALOHA and carrier-sense multiple access with collision avoidance [CSMA/CA]), the secondary users contend for radio resources for data transmission. As in conventional CRNs, the main problem to be addressed in contention-based spectrum access is collision avoidance. However, this problem becomes more complicated due to RF energy harvesting. Firstly, secondary users have to decide whether to harvest RF energy or contend for data transmission. Secondly, to avoid collision, a backoff mechanism can be applied. These decisions must consider the level of remaining energy and amount of RF energy to be harvested. For example, if the channel contention is high, some secondary users should back off their transmissions and harvest RF energy instead. This is not only beneficial for reducing collision, but also to increase the energy level. If the primary user re-occupies the channel, the secondary user may remain in the same channel but switch its mode to harvest RF energy. Note that,  in the RF-powered CRNs,  the secondary user needs to switch a channel  not only when the primary user re-occupies the channel, but also when the secondary user needs to harvest RF energy.

\end{itemize}




\section{Channel Selection in RF-Powered Cognitive Radio Networks}
\label{sec:channelselection}

In the previous section, we have discussed the issues in designing dynamic spectrum access methods for an RF-powered CRN. In this section, for dynamic spectrum access in an RF-powered CRN, we will show how we can formulate the problem of channel selection for a secondary transmitter taking RF energy harvesting into account. We will first describe the system model under consideration and then present a Markov decision problem (MDP) formulation for the  problem. Afterwards, we will present some numerical results obtained by solving the MDP formulation. 

\subsection{System Model}

We consider an RF-powered CRN which consists of $N$ primary users and one secondary user. Each primary user $n$ is allocated with the non-overlapping channel $c_{n}$ for data transmission. Therefore, there are $N$ channels in the RF-powered CRN. All the primary users transmit data on a time slot basis. During each time slot, the channel can be idle or busy (i.e., occupied by the primary user for data transmission). The secondary user is equipped with an RF energy harvester and an energy storage which can store $E$ units of energy. The secondary user can select one of the channels. If the selected channel is busy, the secondary user can harvest energy from the channel. Let $\gamma_{n}$ denote the probability that the secondary user succeeds in harvesting a unit of RF energy from channel $c_{n}$. If the secondary user is in the harvesting area of a primary user $n$, the probability of successful RF energy harvesting is one. This probability can be obtained from an experiment (e.g., as listed in Table~\ref{Data}). The harvested energy is stored in the energy storage. On the other hand, if the selected channel is idle, the secondary user can transmit a packet retrieved from its data queue. The secondary user requires $W$ units of energy for data transmission in a time slot. The probability of a successful packet transmission on channel $c_{n}$ is denoted by $\sigma_{n}$. The probability of a packet arrival for the secondary user in a time slot is denoted by $\alpha$. The arriving packet is buffered in the data queue of the secondary user. The maximum capacity of the data queue is $Q$ packets. We assume that the secondary user has only one wireless interface. Therefore, it cannot transmit data and harvest RF energy at the same time. Also, we assume that the receiver node is always ready to receive the transmitted packet.

The  channel selection policy used by the secondary user  is a mapping from the secondary user's state (i.e., the number of packets in the data queue and the energy level of the energy storage) to the action (i.e., the channel to select). The secondary user does not know the status of the channel  (i.e., whether the channels are idle or busy). In this case, the secondary user selects a channel based on statistical information. This information include the probabilities of a channel to be idle and busy, the probability of successful packet transmission if the channel is idle, and the probability of successful energy harvesting if the channel is busy. After selecting the channel, the secondary user performs spectrum sensing to observe the channel status. If the channel status is busy/idle, the secondary user will harvest RF energy/transmit a packet. 


 
To obtain the optimal channel selection policy, we can formulate the MDP problem and solve it.

\subsection{Optimization Formulation}

The state space of the secondary user is defined by the possible number of packets in the data queue and the energy levels in the energy storage, which are bounded by $Q$ and $E$, respectively. The action space is a set of available channels, which the secondary user can select. Given the channel selected by the secondary user, the following state transitions can happen.
\begin{itemize}
	\item {\em The status of the selected channel is idle.} The transitions depend on the packet arrival probability $\alpha$ and the successful packet transmission probability $\sigma_n$ on selected channel $n$.
	\begin{itemize}
		\item The number of packets increases and the energy level decreases: This happens when a packet arrives and a packet transmission is unsuccessful. 
		\item The number of packets remains the same and the energy level decreases: This happens when a packet arrives and a packet is transmitted successfully or no packet arrives and a packet is transmission is unsuccessful.
		\item The number of packets decreases and the energy level decreases: This happens when no packet arrives and a packet is transmitted successfully.
	\end{itemize}
	\item {\em The status of the selected channel is busy.} The transitions depend on the packet arrival probability $\alpha$ and the successful RF energy harvesting probability $\gamma_n$ on selected channel $n$.
	\begin{itemize}
		\item The number of packets increases and the energy level remains the same: This happens when a packet arrives and RF energy harvesting is unsuccessful. 
		\item The number of packets remains the same and the energy level remains the same: This happens when no packet arrives and RF energy harvesting is unsuccessful.
		\item The number of packets increases and the energy level increases: This happens when a packet arrives and RF energy  harvesting is successful.
		\item The number of packets remains the same and the energy level increases: This happens when no packet arrives and RF energy harvesting is successful.
	\end{itemize}
\end{itemize}
A packet is transmitted successfully if  the energy storage is not empty and there is no wireless channel error. 
Note that the number of packets cannot increase if the data queue is full. Similarly, the energy level cannot increase if the energy queue is full. Conversely, the number of packets cannot decrease if the data queue is empty, and the same is true for the energy level. 
The transition probability matrix for the MDP can be derived according to the above state transitions.

We formulate an optimization problem based on an MDP. We obtain an optimal channel selection policy denoted by $\pi^{\star}$ to maximize the long-term average throughput of the secondary user. The optimization problem is expressed as follows:
\begin{align}
\max_{\pi}: & \quad \mathscr{J}_{T}(\pi) =	\lim_{t \to \infty} \inf\frac{1}{t} 	\sum^{t}_{t^{\prime}=1} \mathbb{E}(\mathscr{T}(\theta_{t^{\prime}}, a_{t^{\prime}}))   
\label{primary} 
\end{align}
where $\mathscr{J}_{T}(\pi)$ is the throughput of the secondary user and $\mathscr{T}(\theta_{t^{\prime}}, a_{t^{\prime}})$ is an immediate throughput function given state $\theta_{t^{\prime}}$ and action $a_{t^{\prime}}$ at time $t^{\prime}$. Note that this optimization problem does not require any constraint for energy harvesting, since if there is not enough energy in the storage, the secondary user cannot transmit a packet and is forced to harvest RF energy. 

Let the state variable be defined as $\theta=(e,q)$ where $e$ and $q$ are the energy level of the energy queue and the number of packets in the data queue, respectively. The immediate throughput function is defined as follows:
\begin{eqnarray}
 \mathscr{T}(\theta, a )	=	\left\{
 \begin{array}{lll}
 \eta_a \sigma_a,       &      & { e \geq W \hspace{2mm}  \text{and}  \hspace{2mm} q>0    } \\
 0,     &      & {\text{otherwise.}}
 \end{array} \right. \end{eqnarray}
where $\theta = (e,q)$ is a combined state of energy level $e$ and number of packets $q$ in the energy storage and data queue, respectively. $\eta_a$ is the probability of the selected channel $a$ (i.e., an action) to be idle and $\sigma_a$ is the probability of successful transmission by the secondary user on the selected channel $a$. In other words, the secondary user successfully transmits a packet if there is enough energy, the queue is not empty, and the selected channel is idle. The packet arrival probability, probability of successful packet transmission, and probability of successful RF energy harvesting determine the state transitions toward condition $e	\geq W$ and $q>0$, which results in immediate throughput. The optimal policy can be obtained by using a standard approach (e.g., through solving an equivalent linear programming problem, value/policy iteration algorithms, and Q-learning algorithm).





\subsection{Performance Evaluation}

\subsubsection{Parameter Setting}

We consider a secondary user whose energy storage and data queue sizes are $10$ packets and $10$ units of energy, respectively. The secondary user requires $1$ unit of energy for $1$ packet transmission. The packet arrival probability is $0.5$. The primary users have two licensed channels $c_{1}$ and $c_{2}$. The probabilities that the channels $c_{1}$ and $c_{2}$ will be idle are $0.1$ and $0.9$, respectively, unless otherwise stated. The probability of successful packet transmission on both channels is $0.95$. The probabilities of successful RF energy harvesting with one unit of energy on channels $c_{1}$ and $c_{2}$ are $0.95$ and $0.70$, respectively, if they are occupied by the primary users. For comparison, we further consider a static policy, in which, a secondary user selects a channel without considering the states of the data queue and energy storage. 

\subsubsection{Numerical Results}

\begin{figure}
\centering
\includegraphics[width=0.5\textwidth]{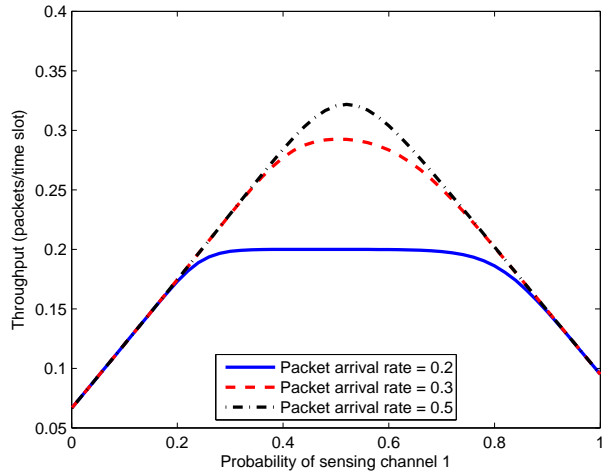}
\caption{Tradeoff of sensing and throughput.} \label{result_mag_static}
\end{figure}

We first show the effect of channel selection (for sensing and subsequently  to transmit a packet or harvest RF energy) on the throughput performance of a secondary user. Figure~\ref{result_mag_static} shows the throughput of the secondary user when the probability of sensing and accessing channel 1 is varied. Interestingly, the throughput can be low if the secondary user senses channel 1 too little or too much frequently. This is from the fact that the channel 1 is mostly occupied by the primary user. Therefore, sensing channel 1 more frequently will result in larger amount of RF energy harvested. However, the secondary user will have less opportunity to transmit packets. On the other hand, sensing channel 1 less frequently will result in smaller amount of RF energy harvested. As a result, the secondary user has a higher chance of having insufficient energy for packet transmission. There could be the optimal ratio that the secondary user should sense and access one particular channel. 

From Fig.~\ref{result_mag_static}, we also observe that the peak throughput at the different packet arrival rate could be different. For example, the peak throughput of the secondary user with packet arrival rate of 0.2 packets/time slot is lower than that with 0.5 packets/time slot. This is due to the fact that the secondary user does not need much energy to transmit packets when the packet arrival rate is small. Therefore, increasing the ratio of sensing and accessing channel 1 does not improve the throughput much.

\begin{figure}
\centering
\includegraphics[width=0.5\textwidth]{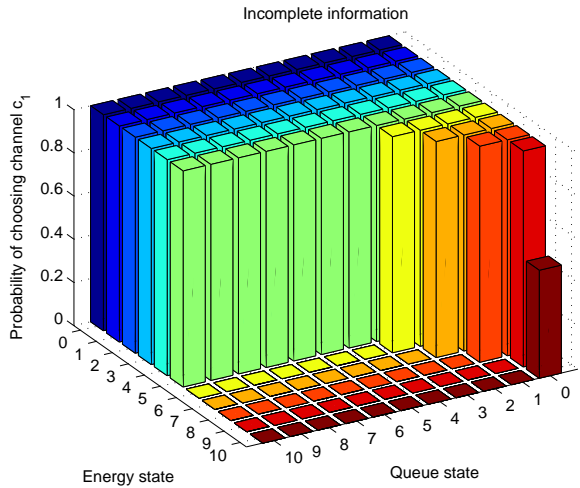}
\caption{Optimal policy of a secondary user.} \label{policy}
\end{figure}

Next, we examine the optimal channel selection policy, where the secondary can optimally choose when to access a particular channel to sense and subsequently access  for packet transmission or use it for RF energy harvesting. Figure~\ref{policy} shows the optimal channel selection policy of the secondary user obtained from the MDP. 
A small probability of selecting channel $c_1$ means the secondary user is likely to select channel $c_{2}$. We observe that the secondary user selects channel $c_{1}$ or $c_{2}$ depending on the data queue and energy storage states (i.e., the energy level and the number of packets, respectively). In this case, the secondary user selects channel $c_{1}$ when the energy level is low and the number of packets in the data queue is small. This is due to the fact that channel $c_{1}$ is more likely to be busy (i.e., available for RF energy harvesting). On the other hand, the secondary user selects channel $c_{2}$ when the number of packets in the data queue is large and energy level is high. This is because  channel $c_{2}$ has higher chance to be idle, which is good for packet transmission by the secondary user. Note that the channel selection policy favors the secondary user to select channel $c_{1}$ more than channel $c_{2}$ since the probability of successful RF energy harvesting from channel $c_{2}$ is lower than that from channel $c_{1}$. 


\begin{figure}
\centering
\includegraphics[width=0.5\textwidth]{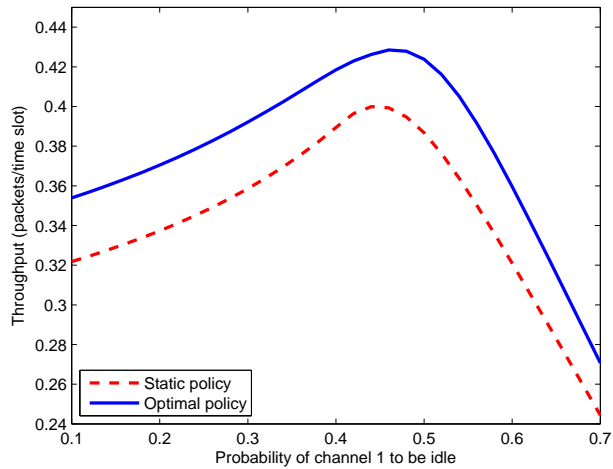}
\caption{Throughput under different dle probability of channel $c_{1}$.} \label{result_mag_optimal}
\end{figure}

We then investigate the case when the idle probability of channel $c_{1}$ is varied (Fig. \ref{result_mag_optimal}). As the idle probability of channel $c_{1}$ increases (i.e., becomes less busy), the throughput of the optimal policy first increases. This is due to the fact that the secondary user has more opportunity to transmit its packets. However, at a certain point, the throughput decreases. This  is due to the fact that when channel $c_{1}$ is mostly idle, the secondary user cannot harvest much RF energy. Therefore, there is not enough energy in the energy storage to transmit packets, thus the throughput decreases. We also provide a comparison with a static policy scheme. For the static policy, the secondary user adjusts the ratio of selecting different channels until the maximum throughput is achieved. In this static policy, the number of packets in the data queue and energy level of the energy storage are not taken into account. Consequently, the throughput of the static policy is lower than that of the optimal policy obtained from the MDP. 

\section{Conclusion}
\label{sec:conclusion}

Opportunistic RF energy harvesting is a promising technique to sustain the operation of secondary users in RF-powered CRNs. In this article, we have discussed the channel selection problem with incomplete information in a CRN which consists of multiple primary users and a secondary user with energy harvesting capability. We have outlined the formulation of an optimization problem based on Markov decision process to obtain the optimal channel selection policy such that the throughput of the secondary user is maximized. We have observed that it is not always beneficial for the secondary user if a channel becomes idle often. In such a scenario, the secondary user cannot harvest enough RF energy from the primary user for its own data transmission resulting in a reduced throughput performance. 
The system model can be extended by considering the channel selection problem in a multi-channel RF-powered CRN composed of multiple secondary users. In this case, the secondary users need to contend for the idle spectrum for transmission under an RF harvesting constraint. The channel selection problem in an RF-powered CRN can be also formulated considering dedicated RF energy sources. In such a scenario, the network allows energy trading between secondary users and dedicated RF energy sources. The secondary users need to decide whether to buy energy from the dedicated RF energy sources or harvest free energy from nearby transmitting primary users.


\end{document}